\documentstyle[prl,multicol,aps,psfig]{revtex}
\begin{document}
\draft
\title{ Scaling in directed dynamical small-world networks with
random responses }
\author{Chen-Ping Zhu$^{1,2}$, Shi-Jie Xiong$^{2}$, Ying-Jie Tian$^{3}$,
Nan Li$^{3}$, Ke-Sheng Jiang$^{3}$}
\address{$^1$Department of Applied Physics, Nanjing University of Aeronautics
and Astronautics, Nanjing, 210016, China\\
$^2$National Laboratory of Solid State Micro-structure and Department of
Physics, Nanjing University, Nanjing 210093, China \\
$^3$College of Economics and Management, Nanjing University of
Aeronautics and Astronautics, Nanjing 210016, China\\}
\date{\today}
\maketitle
\begin {abstract}
    A dynamical model of small-world network, with directed links which
describe various correlations in social and natural
phenomena, is presented. Random response
of every site to the imput message are introduced to simulate real systems.
The interplay of these ingredients results in collective dynamical evolution
of a spin-like variable $S(t)$ of the whole network. In the present
model, global average spreading length $\langle L \rangle_{s}$ and
average spreading time $\langle T\rangle_{s}$ are found to scale
as $p^{-\alpha}ln N$ with different exponents. Meanwhile, $S$ behaves in a 
duple scaling
form for $N\gg N^{*}$: $S\sim f(p^{-\beta}q^{\gamma}\tilde{t})$, where $p$
and $q$ are rewiring and external parameters, $\alpha, \beta,
\gamma $ and $f(\tilde{t})$ are scaling exponents and universal functions,
respectively. Possible applications of the model are discussed.
\end{abstract}
\begin{multicols}{2}
Many natural, social and economic systems,
are well described as networks in which sites represent agents and
links represent correlations between them (see \cite{s1,s2} for
reviews). Among networks with complex topology, small-world
networks (SWN) \cite{s3}
are based on a highly connected regular lattice in which a
fraction $p$ of original links between sites are replaced by
randomly rewired links. It has played an important role in the
study of real processes \cite{s4}, such as various problems in the
field of physics \cite{s5}, and those outside it
\cite{s6,s7,s8,s9,s10}.

As a platform for understanding processes in systems where SWN
topology crucially determines their dynamics, Watts \cite{s4},
Lago-Fern\'andez {\it et al.} \cite{s6}, Strogatz\cite{s9}, and,
Kuperman and Abramson \cite{s10} have analyzed several types of
real problems. Activities of individual agents (sites) in their
models evolve in time, and couple with each other to induce system
response from global coherence, while connections (links) of
networks are kept unchanged throughout the process. However,
interactions in most social or biological systems are not static
and change with time from common knowledge. When states of sites
at the ends of a link change with time, correlation between them
could correspondingly change. This effect can result in novel
dynamical scaling properties which do not appear in previous
works.

Suppose an exclusive shop sells a new brand of portable commodity
without advertising, this message can be spread out by showing it
in the population, and every new buyer can be a source of message,
which change his/her role in the crowd in this sale process.
Meanwhile, interacting relations among people change continuously,
they are not fixed due to everyday activities. Similar things can
occur when a contact-infectious disease with all possible
incubation (infected-but-not-infective) time propagate through an
acquaintance network starting from an index patient. When an
infected one finally becomes infective, he changes his role to all
others. Sometimes people in incubation periods continuing their
social activities even are unaware of infection.

In most information or disease diffusion processes effective
interactions are asymmetric or even unidirected, which has been
investigated in a lot of literature \cite{s11,s12,s13,s14}. The
direction of a link between a pair of given sites depends on their
states. For example, infection passes from infectives to
susceptibles, which defines an active link from the former to the
later. Meanwhile, any link between two infectives or two
susceptibles is inactive in an epidemic spreading. As another
example, some special cases of food webs which show small-world effect
\cite{s15,s16}, can be described with inherent directed trophic links
\cite{s11} and their structures can vary with time \cite{s12}.
These models can illustrate how human pollution or disaster on the
environment to be transported up through species at different
trophic levels.
Recently a non-equilibrium phase transition is displayed in
two-dimensional directed SWNs \cite{s17}.

Usually, individuals in real systems react to input message not
simultaneously \cite{s6,s10}. Incubation time of people for the
same infectious disease varies depending on their resistivity to
it. In parallel, delaying interval from getting a message of the
commodity to buying it also vary according to the spending potential
of a particular buyer. In one word, there always exist stochastic
relaxations or random responses in such dynamical progresses.

In this Letter we attempt to point out that three ingredients
mentioned above---dynamically varied interactions, directionality
of links, and random responses of agents can interweave together in
a real system, and their interplay motivates the present work. We
combine these ingredients to form a dynamical directed small-world
network (DDSWN) on which we simulate practical spreading
processes. With the model dynamical scaling properties of message
diffusion and collective response to the message are revealed by
numerical simulation and data analysis, which have not been
observed from static undirectional SWN models previously.

The network can be built with the following rules: (1) A rewired
one-dimensional (1D) circular lattice with coordination number
$2z$ is constructed. A site labelled $j$ is randomly chosen from
all the $N$ sites as the only initial seed in spin-up state, while
other sites of indices $i\not=j$ are in spin-down state. So the
seed is the first one that is able to send out the message to
other sites and to convert spin states of receiving sites.
(2) Considering the discrepancies among sites in the response to
the message, we define the response time $\tau_{i}$ for every site
$i$ as the relaxation period from the receiving of the message to
 the flipping into the spin-up state. The $\tau_i$'s are random
variables obeying distribution probability $P(\tau_i)$. In this
paper we consider two types of probability, the uniform and
Poisson distributions. The former can be written as
\begin {equation}
P(\tau_{i}) = \frac{1}{q} \theta(\tau_i) \theta(q-\tau_i),
\end {equation}
where $\theta(x)$ is the step function, being 0 for $x<0$ and 1
for $x\geq 0$, and $q$ is a parameter characterizing the width of
the distribution. Here $\tau_i$ is uniformly distributed between
$0$ and $q$, so $q/2$ is the average response period. For the
latter, we use the discrete Poisson distribution
\begin {equation}
P(\tau_{i})=q^{k}e^{-q}/k!,
\end {equation}
where $k$ is the integer part of $\tau_i$, and $q$ corresponds to
the average value of $\tau_{i}$. In both distributions parameter
$q$ measures the average and diversity of the random response
periods. (3) We suppose that the spatial configuration of $\tau_i$
is time independent, but
connections in SWN are updated in every time unit $\tau_0$, i.e.,
a SWN with new links is produced at every moment $t_M = M \tau_0$
with $M$ being a positive integer, from the regular 1D circular
lattice by rewiring $p$ fraction of bonds. So we have dynamically
varied SWNs with the same rewiring probability $p$. (4) All sites
evolve their spin states in a parallel way. Once a site of
spin-down state is at the end of its relaxation time after
receiving the message, it flips onto spin-up state and keeps it,
being capable of sending out the same message. Therefore,
the links in this dynamical SWN are directional as the message can
only be transmitted from a spin-up site to a spin-down site, and
the transmission from a spin-down site or to a spin-up site is
ineffective. Actually, the direction of a link reflects the
irreversible nature in diffusion processes, such as the spreading
of pollution in food webs \cite{s9,s11} and diffusion in directed
percolation models \cite{s18}.

For an undirected and static SWN, the small-world effect can be
described by the characteristic path length $\bar{l}$, i.e., the
averaged shortest distance between any two sites. It follows the
scaling law \cite{s1,s2},
 \begin{equation}
 \label{sca}
\bar{l}(N,p)\sim (N^{*})^{1/d}F(N/N^{*}),
 \end{equation}
where the scaling function $F(u)$ has the limits $ F(u)\sim
u^{1/d}$  for
 $ u \ll 1$, $F(u)\sim \ln u$ for $u\gg 1$, and $N^{*} \sim p^{-1}$ is a
crossover size separating the big- and small-world
regimes. 
In DDSWN, however, not all sites are effectively connected with
each other since only links from sites in spin-up state to those
in spin-down state are active. Therefore, $\bar{l}$ loses its
original meaning in the present model. In order to investigate how
wide and how fast the message spreads out in DDSWNs, we introduce
the time-dependent average spreading length, $\langle
L_{\mu}\rangle \equiv \sum_{\nu=1}^{\mu}{n_{\nu}l_{\nu}}/
\sum_{\nu=1}^{\mu}{n_{\nu}}$, and the average spreading time,
$\langle T_{\mu}\rangle \equiv \sum_{\nu=1}^{\mu}{n_{\nu}t_{\nu}}/
\sum_{\nu=1}^{\mu}{n_{\nu}}$, where $\mu$ is the time index,
$t_{\mu}=\mu t_0$, $n_{\mu}$ is the number of receivers created
between the $(\mu-1)$th and the $\mu$th moments, and $l_{\mu}$ is
the average length from the initial seed to these receivers. We
calculate them by numerical simulations in which the data are
obtained by ensemble average over at least 50 random
configurations with different $\{ \tau_i \}$ and location of the
initial seed. The obtained results are shown in Figs. 1 and 2.
\begin{figure}[h]
\unitlength=1cm
\begin{center}
\begin{picture}(6,5.4)
\put(-2.2,6.6){ \includegraphics{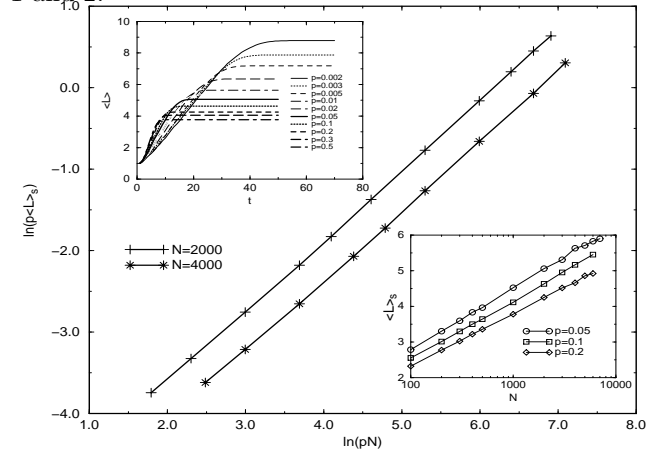}}
\end{picture}
\end{center}
\caption{ Log-log plot of saturated average path length $\langle L
\rangle_{s}$, in the case of uniform distribution of $\tau_{i}$,
versus $p$. Two lines have the same average slope, 0.86.
Upper-left inset: $\langle L \rangle$ as a function of $t$ for the
curve with symbols (+) in the main panel. Lower-right inset:
log-linear relation of $\langle L \rangle_{s}$ versus $N$. Other
parameters for these curves are $z=5$ and $q=10$, and the time is
in units $t_0$. } \label{Fig.1}
\end{figure}
\begin{figure}[h]
\unitlength=1cm
\begin{center}
\begin{picture}(6,4.8)
\put(-2.2,6.6){\includegraphics{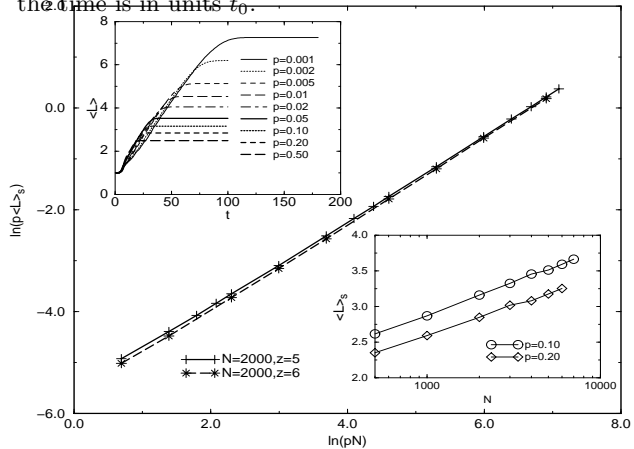}}
\end{picture}
\end{center}
\caption{ The same as those in Fig. 1, but for Poisson
distribution of $\tau_{i}$. In this case the average slope of
lines in the main panel is 0.84. $q=10$ and other parameters are
indicated in the figure. } \label{Fig.2}
\end{figure}

It is found that for given $N$ and $z$ the average path length
$\langle L(t)\rangle$ (here we indicate the time-dependence
explicitly and omit the suffix $\mu$) increases monotonically in
time towards its $p$-dependent saturation value $\langle
L\rangle_{s}$ (see the upper-left insets of Figs. 1 and 2), which
shows the finite-size effect of $p$. Summation over $n_{\nu}$ in
the calculation of $\langle L\rangle_{s}$ tells us that eventually
all sites will be successively visited by the message sooner or
later, suggesting the higher adaptability of the dynamical models
in description of the diffusion processes compared with the static
directed SWN models \cite{s14}. We will use $\langle L\rangle_{s}$
and $\langle T\rangle_{s}$ to characterize global properties of
the networks, since in saturation the quantities are no longer
time dependent. Compared with the scaling relation Eq. (\ref{sca})
for $\bar{l}(N,p)$ in the static SWNs, the corresponding quantity
$\langle L\rangle_{s}$ in the DDSWNs scales with $\ln N$ for given
$p$ (see the lower-right insets of Figs. 1 and 2), but it depends
on $p$ in a power law
 $\langle L\rangle_{s} \sim p^{-\alpha_{L}}$ for a fixed $N$
(see the main panels of Figs. 1 and 2, where the slope of straight
lines gives the value $1-\alpha_{L}$). Therefore, we have a new
scaling formula
\begin {equation}
\langle L \rangle_{s} \sim p^{-\alpha_{L}}\ln N ,
\end {equation}
where $\alpha_{L}=0.14\pm 0.02 $ for uniform distribution and
$\alpha_{L}=0.160\pm 0.005$ for Poisson distribution, extracted
from the data in Figs. 1 and 2, respectively. Meanwhile, the
average spreading time $\langle T \rangle_{s}$ behaves in the same
form as $\langle L \rangle_{s}$, but with different parameters. We
obtain $\alpha_{T} =0.29\pm 0.01 $ and $\alpha_{T} =0.245\pm0.006
$ for uniform and Poisson distributions, as extracted from the
data in Figs. 3 and 4, respectively. The message from the initial
seed passes through tree-like paths, and $\langle L \rangle_{s}$
and $\langle T \rangle_{s}$\\
\begin{figure}[h]
\unitlength=1cm
\begin{center}
\begin{picture}(6,5.8)
\put(-2.2,6.6){ \includegraphics{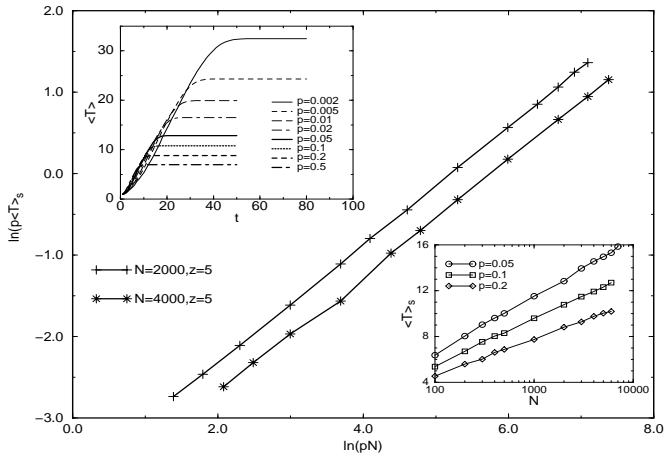}}
\end{picture}
\end{center}
\caption{ Saturated average spreading time $\langle T \rangle_{s}$
as a function of $p$ for the uniform distribution of the response
time. Other parameters are the same as those in Fig. 1. The
average slope of the lines in the main panel is 0.81. }
\label{Fig.3}
\end{figure}
\parindent 0pt are the maximal path length and
spreading time by which the message reaches the final sites. The
scaling relations of $\langle L\rangle _{s}$ and $\langle T
\rangle_{s}$ are unique features of DDSWN model. The simulations
suggest that the $p$-dependence of the saturated $\langle
L\rangle$ and $\langle T \rangle$ is attributed to the dynamically
varied connections and the finite-size effect, the factor $\ln N$
in the scaling relation can be attributed to the geometrical
characteristics of the SWN background, and the factor
$p^{-\alpha_{L}}$ is originated from the diffusion mechanism of
the directional links, distinctive in comparison with the behavior
of $\bar{l}$ in the ordinary SWNs. The exponents $\alpha_{L}$ and
$\alpha_{T}$ still depend on specific form of the distributions of
the relaxation time.
On the other hand,
as it should be anticipated, $\langle L\rangle _{s}\sim N$ for $N
\ll N^{*}$ in the big-world regime.

\begin{figure}[h]
\unitlength=1cm
\begin{center}
\begin{picture}(6,5.8)
\put(-2.2,6.6){ \includegraphics{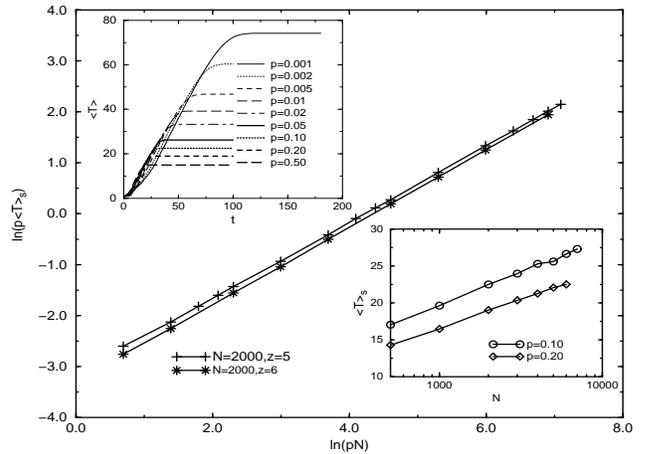}}
\end{picture}
\end{center}
\caption{ The same as those in Fig. 3, but for the Poisson
distribution of the response time. Other parameters are the same
as those in Fig. 2. The average slope of the lines in main panel
is 0.755. } \label{Fig.4}
\end{figure}

\parindent 30pt We now turn to the behavior of $S(t)$, defined as the number of
all sites in spin-up states at moment $t$ divided by $N$. It
represents the ratio of the sites in spin-up state, and its
temporal derivation, $S'(t)$, reflects the probability of a
spin-down site to be flipped into the spin-up state in a unit of
time at moment $t$. Because in this model all the sites should be
in spin-up state after an enough time, $S(t)\rightarrow 1$ at the
limit $t\rightarrow \infty$. We numerically calculate $S(t)$ for
various values of parameters. The results for different values of
$p$ and fixed $q$ in the case of uniform distribution are plotted
in Fig. 5(a). All the curves can merge together if the time is
rescaled as $t_{\text{sc}} = (p/p_0)^{\beta}t$, as can be seen
from Fig. 5(b) in which we use $p_0=0.1$ and $\beta=0.20\pm 0.005$
is extracted. The time rescaling is done for the Poisson
distribution, as shown in Figs. 5(c) and 5(d), and we get $\beta
=0.186 \pm 0.006$. At the same time, if $q$ is varied as well, we can
also use the time rescaling, $t_{\text{sc}}' = (q/q_0)^{-\gamma}
t_{\text{sc}} $ to merge the curves, as shown in Fig. 6. From this
we extract the exponent $\gamma = 0.750 \pm 0.003$ for the uniform
distribution and $\gamma = 0.880 \pm 0.006$ for the Poisson
distribution, both with $q_0=10$. Thus, we have the following scaling 
form of $S(t)$:
 \begin {equation}
S(t) \sim S(p^{-{\beta}}q^{\gamma}\tilde{t})\equiv f(\tilde{t}),
 \end {equation}
where $\tilde{t} = {p_0}^{-\beta}{q_0}^{\gamma}t'_{\text{sc}}$ and
$f(\tilde{t})$ is
a universal function for a given distribution.

\begin{figure}[h]
\unitlength=1cm
\begin{center}
\begin{picture}(6,5.8)
\put(-2.2,6.6){ \includegraphics{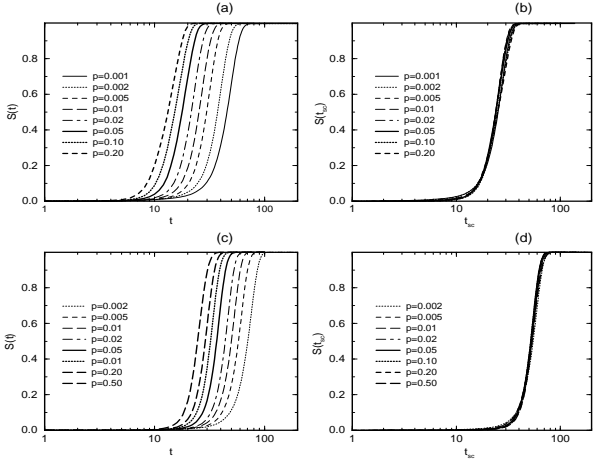}}
\end{picture}
\end{center}
\caption{ The ratio of spin-up sites versus time $t$ (a and c) and
versus rescaled time $t_{\text{sc}}= (p/p_0)^{\beta} t$ (b and d)
for the uniform (a and b) and Poisson (c and d) distributions.
$p_0 =0.1$ is used. Other parameters are: $N=2000$, $z=5$, and
$q=10$. } \label{Fig.5}
\end{figure}

\begin{figure}[h]
\unitlength=1cm
\begin{center}
\begin{picture}(6,5.8)
\put(-2.2,6.6){ \includegraphics{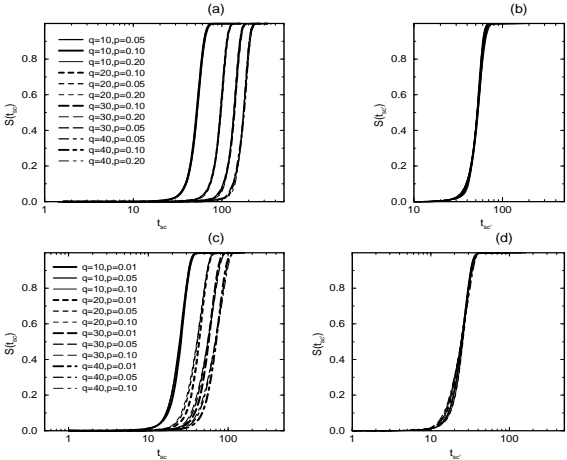}}
\end{picture}
\end{center}
\caption{ The ratio of spin-up sites versus $t_{\text{sc}}$ (a and
c) and versus $t_{\text{sc}}'= (q/q_0)^{-\gamma } t_{\text{sc}}$
(b and d) for Poisson (a and b) and the uniform (c and d)
distributions. $q_0=10$ is used. Other parameters are: $N=2000$ and $z=5$. }
\label{Fig.6}
\end{figure}

The distributions of the response time used here are
phenomenological functions reflecting the nature of the DDSWN.
Other forms can also be used to describe specific systems. The
results of our simulations with different distributions suggest
that there exists common scaling behavior of the diffusion
dynamics in DDSWNs and only the values of exponents depend on the
form of distribution. For a specific process, the response time
has a definite meaning. Sales problem of a new type of commodity,
for example, $\tau_{i}$ reflects the delaying time between the
awareness and the buying of the commodity for a consumer, and $q$
is related to the discrepancy in consumptive potential among them.
Food webs, as another example, usually have a few directed links
\cite{s16} from the basal species to the top ones, and moreover, have
average trophic distance scaling in $\ln N$\cite{s19}.
All of them share typical values of $\langle L \rangle_{s}$ in
the DDSWN model.
This implies that the model can capture the essential feature of
dynamical directionality of food webs although the small-world
effect in this problem is still in dispute \cite{s15,s16}. By the
way, on revision of our submitted manuscript, similar $\ln N$
scaling behaviors of SWNs were independently found in quantum and
classical diffusion problems \cite{s20}.

This work is supported
by China Aeronautic Science Fundation under Grant No.00J52079.

\end{multicols}
\end{document}